\documentclass[pdflatex,sn-aps]{sn-jnl}
\usepackage{graphicx}%
\usepackage{multirow}%
\usepackage{amsmath,amssymb,amsfonts}%
\usepackage{amsthm}%
\usepackage{mathrsfs}%
\usepackage[title]{appendix}%
\usepackage{xcolor}%
\usepackage{textcomp}%
\usepackage{manyfoot}%
\usepackage{booktabs}%
\usepackage{algorithm}%
\usepackage{algorithmicx}%
\usepackage{algpseudocode}%
\usepackage{listings}%
\usepackage{float}%
\usepackage{enumitem}%
\usepackage[separate-uncertainty=true]{siunitx}%

\begin{document}

\title{Windowing in terahertz time-domain spectroscopy: resolving resonances in thin-film samples}

\author[]{\fnm{Esteban} \sur{Marulanda}}

\author[]{\fnm{Fernanda L.} \sur{Costa}}

\author[]{\fnm{Nicolas M.} \sur{Kawahala}}

\author*[]{\fnm{Felix G. G.} \sur{Hernandez}}\email{felixggh@if.usp.br}

\affil[]{\orgdiv{Instituto de Física}, \orgname{Universidade de São Paulo}, \orgaddress{\city{São Paulo}, \postcode{05508-090}, \state{SP}, \country{Brazil}}}

\abstract{Terahertz time-domain spectroscopy (THz-TDS) has become a powerful tool for investigating the optical properties of thin films, offering direct access to the complex permittivity in the terahertz range. However, in transmission-based measurements of thin films on thick substrates, multiple reflections and limited time windows can introduce artifacts that obscure resonant features such as phonon modes. Time-domain windowing remains one of the most widely adopted strategies to mitigate these effects, yet systematic guidelines on its application remain scarce. In this work, we organize a practical routine for extracting the complex permittivity from THz-TDS data, focusing on when and how to apply time-domain windowing. The routine incorporates decision points for truncation versus smooth apodization, and emphasizes tailoring the window length and shape to specific signal conditions. We demonstrate the approach using representative measurements on PbTe thin films, highlighting cases in which truncation suffices, where apodization is essential, and how different window functions and lengths influence the resulting spectra. We also propose simple metrics to assess signal continuity and guide window selection. Although other analysis techniques exist, including parametric spectral estimation, this study focuses on formalizing windowing-based processing into an accessible experimental workflow. Our results show that the choice of window parameters can significantly affect the accuracy of extracted material parameters, particularly for sharply resonant systems. This work provides an accessible framework for improving spectral fidelity in THz-TDS of thin-film samples.}


\maketitle

\section{Introduction}\label{sec:intro}
Terahertz time-domain spectroscopy (THz-TDS) has become a cornerstone technique for the non-destructive characterization of materials in the terahertz (THz) frequency range \cite{nuss_terahertz_1998,hangyo_terahertz_2005,koch_terahertz_2023}. By recording both the amplitude and phase of the transmitted or reflected THz electric field, THz-TDS enables direct extraction of the complex refractive index and permittivity without resorting to Kramers--Kronig transformations \cite{schmuttenmaer_exploring_2004,lloyd-hughes_review_2012}. This approach has proven effective for investigating diverse systems, including semiconductors \cite{Yan:15,okamura2022terahertz}, superconductors \cite{Lupi2011-mq,WILKE20002271}, topological insulators \cite{top_ins,top2}, van der Waals materials \cite{kawahala_shaping_2025,mitra_terahertz_2025}, pharmaceutical compounds \cite{patil_terahertz_2022,huang_progress_2023}, and biomolecules \cite{doi:10.1021/acs.jpcb.7b02724,penkov_terahertz_2023}.

Spectral analysis in THz-TDS is most commonly performed using Fourier transforms due to their straightforward implementation and direct connection to the time-domain nature of the measurement. However, Fourier-based analysis has well-known limitations, such as the dependence of the frequency resolution on the temporal window length, and the potential for spectral leakage due to discontinuities at window edges. These constraints have motivated the use of complementary spectral estimation methods, including autoregressive (AR), autoregressive moving average (ARMA), and transfer-function-based parametric models \cite{stoica2005, sanjuan2015, tych2014}, wavelet-based techniques for echo separation, and analytical modeling approaches that explicitly account for Fabry–Pérot reflections \cite{duvillaret1999, duvillaret1996, dorney2001, pupeza2007, coutaz2018, labbe-lavigne1998}. Regardless of this diversity of tools, direct Fourier-based methods remain widely adopted due to their accessibility and experimental practicality. Within this context, signal processing steps such as zero-padding and windowing \cite{donnelle_fast_2005,prabhu_window_2014} are frequently applied to suppress artifacts from finite measurement duration \cite{withayachumnankul_fundamentals_2014}. For instance, rectangular windows are routinely used to eliminate internal reflections (Fabry--Pérot effects), while smooth apodization functions reduce spectral leakage from abrupt truncations \cite{harris_use_1978}. However, these operations are often applied empirically, without rigorous evaluation of their impact on THz-TDS analysis.

Despite their widespread use, the effects of windowing on optical property extraction are rarely examined in depth. To our knowledge, only Vázquez-Cabo \textit{et al.} \cite{vazquez-cabo_windowing_2016} have systematically studied window functions in THz spectroscopy, though their work focused on lactose powders. In contrast, multi-layered systems such as thin films---where resonances often overlap with substrate echoes or electronic backgrounds---windowing choices can significantly distort or obscure spectroscopic features.

Thin films present unique challenges and opportunities for THz-TDS \cite{ohara_review_2012}. Their reduced thickness enables transmission measurements even in highly absorptive materials \cite{neu_tutorial_2018}, facilitating studies of phonon modes, carrier dynamics, and phase transitions \cite{hernandez_observation_2023,stefanato_metallicity-driven_2025}. Yet, the resulting signals often exhibit low amplitude and long-lived oscillations, which complicate time-domain separation from substrate artifacts. Consequently, resonance retrieval becomes highly sensitive to windowing parameters (e.g., shape, width, and temporal alignment), but systematic guidelines remain scarce.

Here, we formalize a structured routine for extracting the complex permittivity of thin films from THz-TDS transmission data, incorporating time-domain windowing to minimize artifacts and increase spectral fidelity on widely used Fourier-based approaches. Although such windowing is frequently applied in practice, it is often done heuristically, with limited discussion in the literature about when and how to apply it most effectively. Our aim is to help bridge this gap by consolidating windowing strategies into a coherent, experimentally grounded workflow. To demonstrate its applicability across representative spectroscopic scenarios, we apply it to previously reported THz-TDS data of lead telluride (PbTe) thin films \cite{baydin_magnetic_2022,kawahala_thickness-dependent_2023}, which exhibit well-characterized phonon resonances in the THz range. By comparing different windowing strategies, we identify conditions under which simple truncation enables accurate recovery of spectroscopy features, and when further apodization may be necessary.

\section{Methods}
\subsection{Terahertz time-domain spectroscopy}
THz-TDS is a spectroscopic technique that probes the interaction of materials with broadband, sub-picosecond pulses of terahertz radiation \cite{jepsen_terahertz_2011,koch_terahertz_2023}. In a typical experimental geometry, the time-dependent electric field $E(t)$ transmitted through (or reflected from) a sample is measured directly. Fourier transformation of this time-domain signal yields the complex frequency-domain spectrum $E(\nu)=\mathcal{F}\{E(t)\}$, where $\nu$ denotes the linear frequency, providing simultaneous access to both amplitude and phase without Kramers--Kronig analysis \cite{schmuttenmaer_exploring_2004}.

\begin{figure}[htb]
    \centering
    \includegraphics[width=1\textwidth]{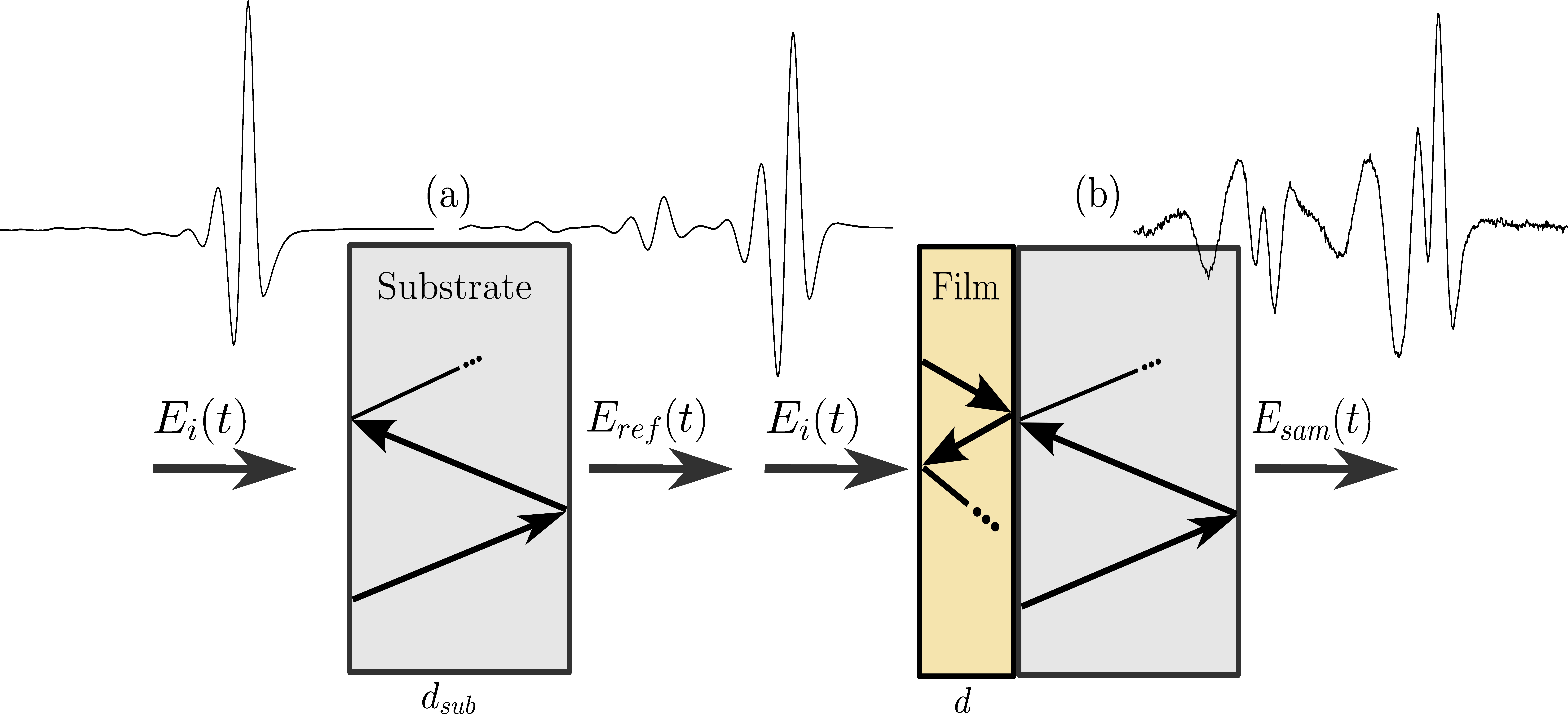}
    \caption{Schematic of THz-TDS measurements for extracting the complex permittivity of a thin film. \textbf{(a)} Reference measurement through a bare substrate of thickness $d_\textrm{sub}$, where the incident THz electric field $E_\textrm{i}(t)$ is reshaped into the reference signal $E_\textrm{ref}(t)$ after transmission. \textbf{(b)} Sample measurement through a thin film of thickness $d$ deposited on an identical substrate, yielding the transmitted sample signal $E_\textrm{sam}(t)$. Multiple internal reflections within the substrate and film layers generate Fabry--Pérot echoes in both waveforms, with pronounced oscillations in the sample due to resonant absorption.}
    \label{fig1}
\end{figure}

For thin films deposited on thick substrates, isolating the film's contribution requires comparing the sample response (film + substrate) with a bare substrate reference, as illustrated in Fig.~\ref{fig1}. In a transmission geometry, this is accomplished by computing the complex transmission coefficient
\begin{equation}\label{eq:trans_coef}
    \mathcal{T}(\nu) = \frac{E_\textrm{sam}(\nu)}{E_\textrm{ref}(\nu)},
\end{equation}
where $E_\textrm{sam}(\nu)$ and $E_\textrm{ref}(\nu)$ are the Fourier-transformed THz electric fields transmitted through the sample and reference substrate, respectively. The transmittance is then $|\mathcal{T}(\nu)|^2$.

To extract optical coefficients, transmission models based on electromagnetic transfer functions are employed. For thin films on thick substrates, the widely used model assumes sufficiently small phase shift and moderate absorption in the film, yielding \cite{lloyd-hughes_review_2012,neu_tutorial_2018}:
\begin{equation}\label{eq:thinFilmModel}
    \mathcal{T}_\textrm{film}(\nu) = \frac{1+n_\textrm{sub}(\nu)}{1+n_\textrm{sub}(\nu)-2\pi i[\epsilon_\textrm{film}(\nu)+n_\textrm{sub}(\nu)]\nu d/c},
\end{equation}
where $n_\textrm{sub}(\nu)$ is the substrate refractive index, $\epsilon_\textrm{film}(\nu)=\epsilon_1(\nu)+i\epsilon_2(\nu)$ is the film's complex permittivity, $d$ is the film thickness, and $c$ is the speed of light in vacuum. Under these conditions, $\epsilon_\textrm{film}(\nu)$ can be obtained by analytically inverting equation~\eqref{eq:thinFilmModel}.

\subsection{Analysis routine for thin-films}
For THz-TDS experiments in transmission geometry involving thin films on thick substrates, we outline a structured routine to assist in extracting the film’s complex permittivity while preserving key physical features and mitigating artifacts. The procedure is particularly useful under limited time-window conditions, where Fabry–Pérot reflections may overlap with the main pulse, distorting spectra. While the individual techniques involved are well-established, our aim is to organize their application into a coherent framework, incorporating qualitative criteria for assessing windowing suitability and practical guidance for implementation. The steps are outlined as follows:

\begin{enumerate}
    \item \textbf{Acquire time-domain signals:} Measure the THz-TDS transmission signals for the substrate-only reference $E_\textrm{ref}(t)$ and for the sample $E_\textrm{sam}(t)$, using a uniform time step $\delta t$. Each signal contains $N$ points, covering a total time window $\tau=N\delta t$. Ensure that the time origin $t_0$ is consistently defined across measurements to avoid artificial phase errors in the frequency domain.
    
    \item \textbf{Determine relative delay:} Identify the time positions of the main pulse maxima in each signal, denoted $t_\textrm{ref}^\textrm{max}$ and $t_\textrm{sam}^\textrm{max}$, and compute the delay $\Delta t=t_\textrm{sam}^\textrm{max}-t_\textrm{ref}^\textrm{max}$. If non-negligible, this delay often results from a small thickness mismatch $\Delta L$ between substrates, which may cause a relative shift of a few picoseconds. 

    \item \textbf{Locate internal reflections}: Identify the onset time of the first internal reflection in each signal as $t_\textrm{ref}^\textrm{refl}$ and $t_\textrm{sam}^\textrm{refl}$.

    \item \textbf{Assess truncation feasibility:} Inspect a short interval (e.g. a few picoseconds) preceding each $t^\textrm{refl}$. Evaluate whether the signal smoothly decays toward baseline or exhibits residual oscillations or abrupt changes:

    \begin{enumerate}[label=\theenumi\alph*.]
        \item if both signals decay smoothly, truncate them as: 
        \begin{equation}
        \begin{aligned}
            \bar{E}_\textrm{ref}(t) &= E_\textrm{ref}(t), \quad t < t_\textrm{ref}^\textrm{refl} \\
            \bar{E}_\textrm{sam}(t) &= E_\textrm{sam}(t), \quad t < t_\textrm{sam}^\textrm{refl}
        \end{aligned}
        \end{equation}
        which have the lengths $N_\textrm{ref}=(t_\textrm{ref}^\textrm{refl}-t_0)/\delta t$ and $N_\textrm{sam}=(t_\textrm{sam}^\textrm{refl}-t_0)/\delta t$, respectively. Match lengths by padding zeros to the shorter signal so both have $N_\textrm{trunc}=\max(N_\textrm{ref}, N_\textrm{sam})$. Jump to Step 8.

        \item Otherwise, if either signal shows non-negligible slope or oscillations near $t^\textrm{refl}$, proceed to the next step.
    \end{enumerate}

    \item \textbf{Align sample to reference:} Shift the untruncated sample signal $E_\textrm{sam}(t)$ by $\Delta t$ to align it with the reference:
    \begin{equation}
        E_\textrm{sam}'(t) = E_\textrm{sam}(t+\Delta t).
    \end{equation}
    Implement this shift numerically by removing $\Delta t/\delta t$ points from the appropriated edge (leading or trailing), and padding zeros on the opposite side, depending on the sign of $\Delta t$.

    \item \textbf{Choose an apodization window:} Select a window function $W(t)$ of length $N_\textrm{win}$. For example, Gaussian windows provide optimal time-frequency resolution trade-offs, while Flattop windows minimize amplitude errors in spectral peaks \cite{vazquez-cabo_windowing_2016,prabhu_window_2014}. The window should taper smoothly to zero at its edges to suppress spectral leakage. Choose $N_\textrm{win}$ carefully: too short, and relevant dynamics may be lost; too long, and internal reflections may leak in unless attenuated by the window tail.

    \item \textbf{Center and apply the window:} Center both $E_\textrm{ref}(t)$ and $E_\textrm{sam}'(t)$ within the apodization window by aligning their main peaks with the window center. This may require trimming and zero-padding as needed. Apply the window to obtain:
    \begin{equation}
        \tilde{E}_\textrm{ref}(t)=E_\textrm{ref}(t)\,W(t), \quad \tilde{E}_\textrm{sam}(t)=E_\textrm{sam}'(t)\,W(t).
    \end{equation}
    Both windowed signals now have length $N_\textrm{win}$.

    \item \textbf{Compute the Fourier transforms:} Perform the discrete Fourier transforms of the processed reference and sample signals as follows. If truncation was applied (Step 4a),
    \begin{equation}
        E_\textrm{ref}(\nu) = \mathcal{F}\{\bar{E}_\textrm{ref}(t)\}, \quad E_\textrm{sam}(\nu) = \mathcal{F}\{\bar{E}_\textrm{sam}(t)\}.
    \end{equation}
    If apodization windowing was used (Step 7),
    \begin{equation}
        E_\textrm{ref}(\nu) = \mathcal{F}\{\tilde{E}_\textrm{ref}(t)\}, \quad E_\textrm{sam}(\nu) = \mathcal{F}\{\tilde{E}_\textrm{sam}(t)\}e^{2\pi i\nu\Delta t},
    \end{equation}
    where the exponential factor corrects the phase shift introduced by the alignment in Step 5. Zero-padding (e.g., extending signals with trailing zeros to a greater power of two) may be applied to improve frequency resolution and enable smoother spectral interpolation, without altering the underlying signal information.

    \item \textbf{Extract the film permittivity:} Compute the complex transmission coefficient, corrected for the substrate thickness difference $\Delta L$ as discussed in Step 2:
    \begin{equation}\label{eq:trans_coef_DeltaL}
        \mathcal{T}_\textrm{film}(\nu) = \frac{E_\textrm{sam}(\nu)}{E_\textrm{ref}(\nu)}e^{2\pi i\nu [1-n_\textrm{sub}(\nu)]\Delta L/c},
    \end{equation}
    using prior knowledge of the substrate refractive index spectrum $n_\textrm{sub}(\nu)$. Then, retrieve the film's complex permittivity by inverting the thin-film model in equation~\eqref{eq:thinFilmModel}:
    \begin{equation}\label{eq:eps_film}
        \epsilon_\textrm{film}(\nu) = i[1+n_\textrm{sub}(\nu)]\left[\frac{1}{\mathcal{T}_\textrm{film}(\nu)}-1\right]\frac{c}{2\pi\nu d} - n_\textrm{sub}(\nu),
    \end{equation}
    where $d$ is the film thickness. If unphysical results occur (e.g., negative $\epsilon_2$ values), verify factors such as window alignment, and $\Delta L$ correction.

\end{enumerate}

\section{Results}
To demonstrate the applicability of the routine, we analyzed THz-TDS data from a thin-film sample grown on a thick substrate, selected as a representative system extensively studied in recent terahertz spectroscopy literature for its characteristic phononic and electronic responses \cite{baydin_magnetic_2022,kawahala_thickness-dependent_2023}. Measurements using a conventional transmission-geometry THz-TDS setup (see Ref.~\cite{kawahala_thickness-dependent_2023}) illustrate the routine's performance across four critical scenarios: Fabry--Pérot suppression via rectangular windowing; discontinuity mitigation through apodization; optimization of window size for resolution-artifact balance; and spectral consequences of window function selection.

\subsection{Fabry--Pérot suppression via rectangular windowing}
As a first application of the compiled routine, we examine the optimal case for Fabry--Pérot suppression through simple time-domain truncation, which acts as a rectangular window that cleanly isolates the main transmitted pulse from well-separated echoes. This avoids spectral artifacts characteristic of internal reflections in thick substrates. To demonstrate, we use representative THz-TDS transmission data from a $\qty{1.6}{\um}$-thick PbTe thin film grown on a bulk BaF$_2$ substrate, measured at a temperature of \qty{120}{\K}, where we expect a well-defined phonon resonance in the sub-THz range ($\nu<\qty{1}{\THz}$) \cite{burkhard_submillimeter_1977,kawahala_thickness-dependent_2023}.

\begin{figure}[t]
    \centering
    \includegraphics[width=1\textwidth]{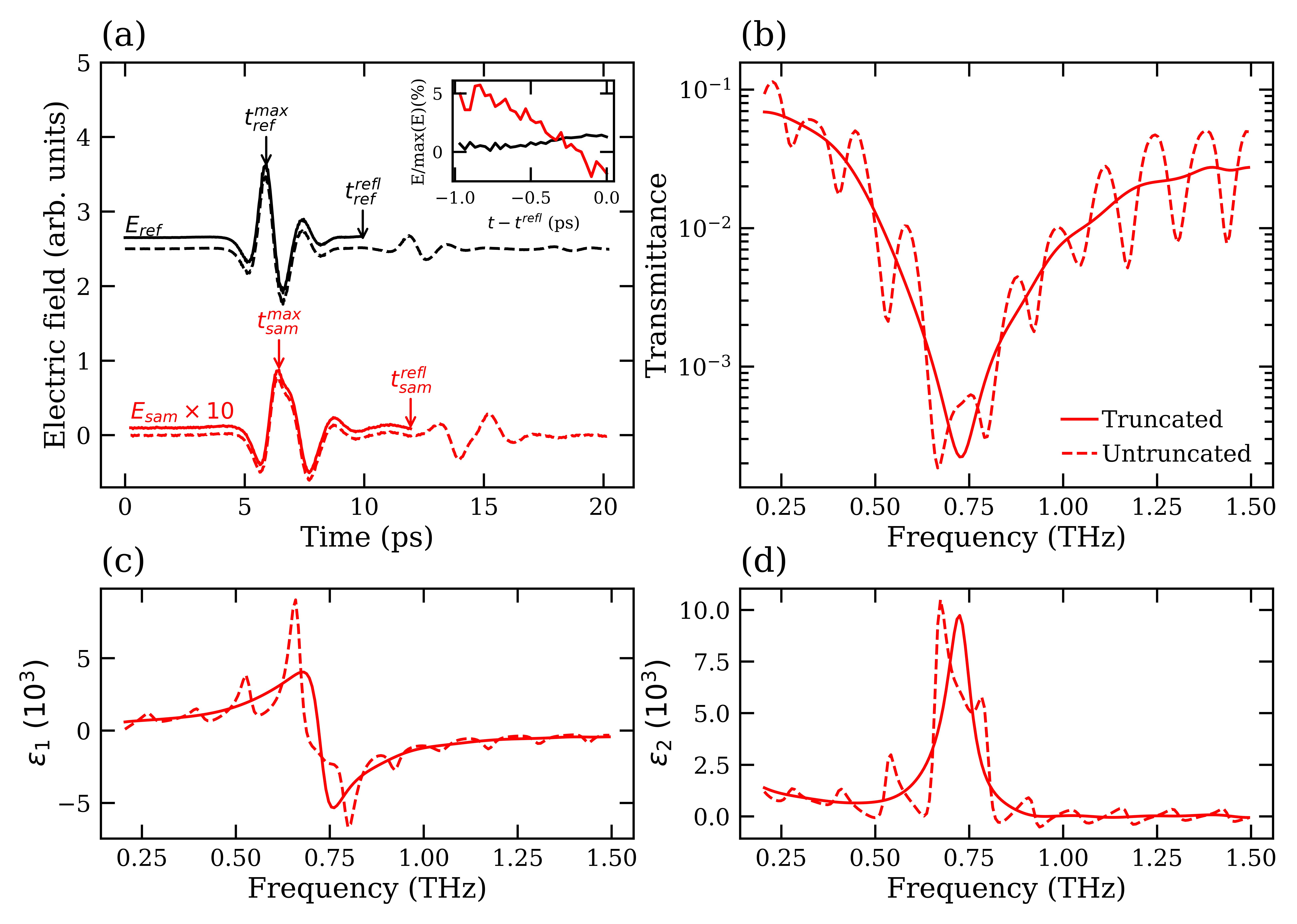}
    \caption{Effect of time-domain truncation on extracted optical properties of a PbTe thin film at \qty{120}{\K}. \textbf{(a)} Time-domain electric field signals for the reference $E_\textrm{ref}(t)$ (black) and sample $E_\textrm{sam}(t)$ (red). For clarity, all curves were vertically offset and the sample waveform is scaled $\times 10$. Arrows mark truncation points $t^\textrm{refl}$ and main peaks $t^\textrm{max}$. Dashed: original signals. Solid: truncated signals. Inset: final \qty{1}{ps} of truncated waveforms, normalized by their respective peak amplitudes, showing smooth decay. \textbf{(b)} Transmittance spectra calculated with (solid) and without (dashed) time-domain truncation. \textbf{(c)} Real and \textbf{(d)} imaginary parts of the extracted complex permittivity of the film. Truncation suppresses Fabry--Pérot oscillations and improves the clarity of the phonon resonance near \qty{0.75}{\THz}.}
    \label{fig2}
\end{figure}

Fig.~\ref{fig2}a shows the time-domain electric field waveforms for the reference ($E_\textrm{ref}$, black) and sample ($E_\textrm{sam}$, red), plotted over the full temporal acquisition range (dashed lines). For a better visualization, the sample signal has been scaled by a factor of \num{10}. Following Step 1 of the routine, both waveforms were acquired with a uniform time step $\delta t=\qty{0.0333}{\ps}$ over $N=\num{600}$ points (total time window of $\tau=\qty{20}{\ps}$). Step 2 analysis reveals a $\Delta t=\qty{0.533}{\ps}$ delay between pulse maxima ($t^\textrm{max}$ arrows), which may be largely attributed to a substrate thickness difference of $\Delta L = \qty{0.086(1)}{\mm}$, measured using a micrometer.

Following Step 3, the onset times of the first internal reflections ($t^\textrm{refl}$) are also marked with arrows in Fig.~\ref{fig2}a. The solid curves depict the truncated signals, where both waveforms were cut immediately before each respective reflection point. The inset zooms into the final \qty{1}{\ps} of the truncated waveforms, normalized to their peak electric field amplitude. The reference signal shows minimal residual content ($\bar{E}_\textrm{ref}/\max(\bar{E}_\textrm{ref})$ increases from \qty{0}{\percent} to \qty{1}{\percent}), while the sample signal decays smoothly from about \qty{5}{\percent} to near zero. This confirms the signals meet Step 4a's truncation criteria: no relevant discontinuities are introduced, and Fabry--Pérot artifacts are effectively suppressed.

Following Step 4a, we proceed to Step 8 for Fourier transform computation of the truncated signals. To achieve enhanced frequency resolution, zero-padding extends each waveform to $2^{12}$ points. Using the substrate refractive index $n_\textrm{sub}\approx\num{2.6}$ (determined from independent THz-TDS measurements on a bare BaF$_2$ substrate, and consistent with literature values \cite{bosomworth_far-infrared_1967}), Step 9 yields the film's complex transmission coefficient $\mathcal{T}_\textrm{film}(\nu)$. For comparison, we analyze both truncated and untruncated signals, with Fig.~\ref{fig2}b showing their transmittance spectra (solid and dashed lines, respectively).

The truncated-data transmittance exhibits characteristic PbTe thin-film behavior at \qty{120}{\K}, showing a broad absorption band centered near \qty{0.75}{\THz} attributed to a transverse optical (TO) phonon mode \cite{jiang_origin_2016,baydin_magnetic_2022}. This physically meaningful response contrasts sharply with the untruncated data, where Fabry--Pérot reflections introduce periodic spectral modulations that mix with the intrinsic phonon feature. These artifacts propagate to the extracted complex permittivity (Figs.~\ref{fig2}c--d): while the truncated data reveals the expected Lorentzian phonon response \cite{fox_optical_2010}, the untruncated analysis yields unphysical oscillations in both $\epsilon_1$ and $\epsilon_2$, confirming the windowing routine's critical role in isolating intrinsic film properties. Although fitting to a Lorentzian model and extracting resonance parameters with statistical confidence is beyond the scope of this work, we estimate $\epsilon_2$ peak from the truncated data as \num{9.7(1)e3} near \qty{0.73(4)}{\THz}. This is in reasonable agreement with literature values for PbTe thin films (e.g., real conductivity peaks in Ref. \cite{baydin_magnetic_2022}).

\subsection{Discontinuity suppression via apodization}
This second case addresses scenarios where simple truncation introduces problematic edge discontinuities, requiring apodization windowing (Steps 5--7 of the routine). As a first illustrative example, we apply a Gaussian window, which is commonly used in THz-TDS due to its smooth temporal profile and frequently cited in introductory references \cite{neu_tutorial_2018}. A broader comparison of window types is presented in Section~3.4. We examine again THz-TDS data from the same PbTe thin film, now measured at \qty{10}{\K}. At this lower temperature, red-shifted enhanced phonon absorption produces a more attenuated transmitted signal with longer-lived THz field oscillations due to the reduced phonon damping \cite{kawahala_thickness-dependent_2023}. These sustained oscillations significantly increase the risk of slope discontinuities upon truncation, necessitating smooth windowing to maintain spectral accuracy.

\begin{figure}[t]
    \centering
    \includegraphics[width=1\textwidth]{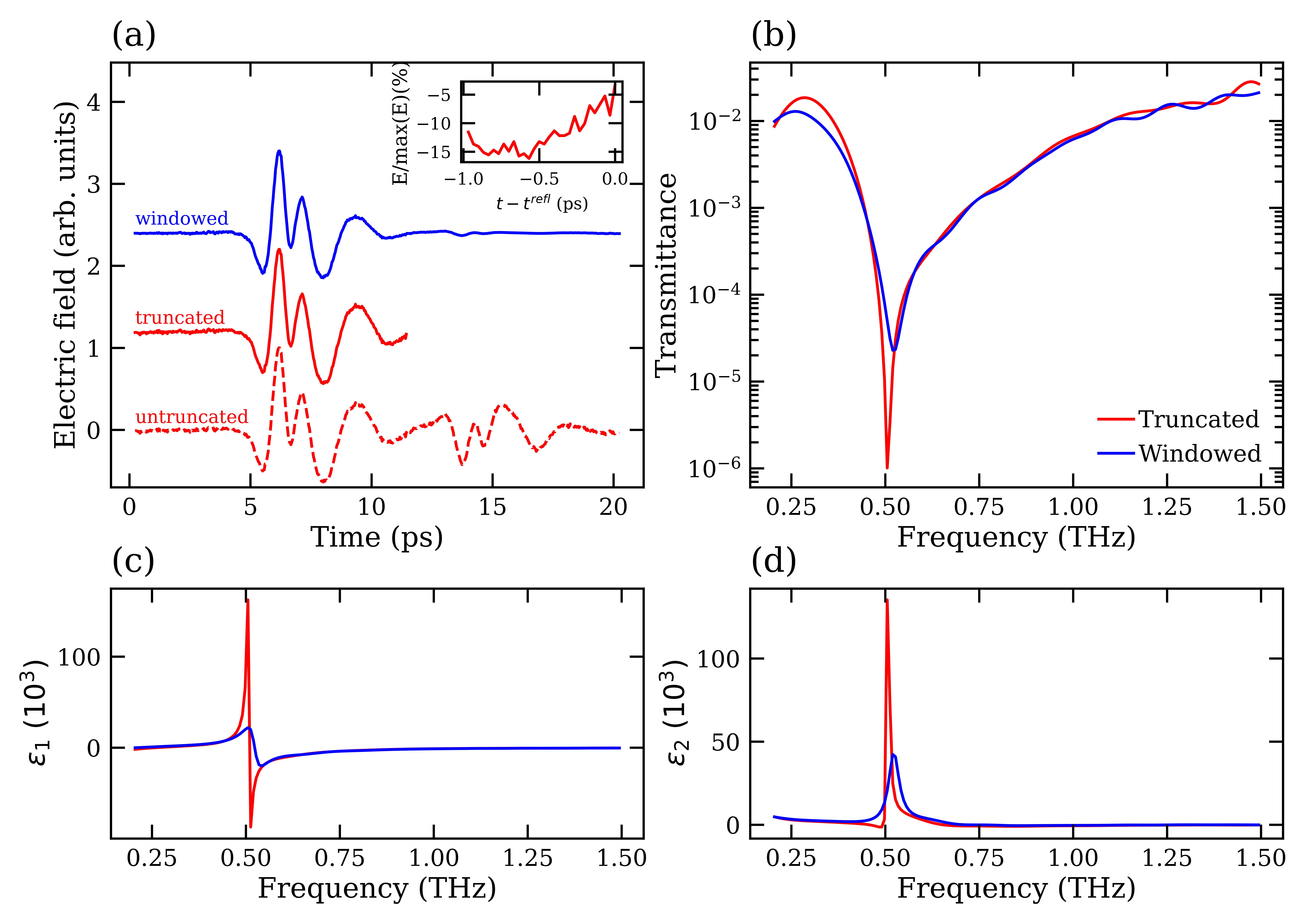}
    \caption{Effect of apodization windowing on extracted optical properties of a PbTe thin film at \qty{10}{\K}. \textbf{(a)} Sample time-domain waveform $E_\textrm{sam}(t)$ under three processing conditions: untruncated (dashed red), truncated (solid red), and Gaussian-windowed (solid blue). For clarity, all curves are vertically offset. Truncation suppresses internal reflections but introduces a sharp discontinuity. Inset: final \qty{1}{\ps} of truncated waveform, normalized by its peak amplitude, showing rapid slope change exceeding \qty{10}{\percent} of peak value. \textbf{(b)} Transmittance spectra calculated from truncated (solid red) and Gaussian-windowed (solid blue) signals. \textbf{(c)} Real and \textbf{(d)} imaginary parts of the extracted complex permittivity of the film. Gaussian windowing suppresses spectral artifacts and yields a more physically realistic resonance near \qty{0.5}{\THz}.}
    \label{fig3}
\end{figure}

To demonstrate, Fig.~\ref{fig3}a compares the sample time-domain waveform under three processing conditions: untruncated (dashed red), truncated (solid red), and Gaussian-windowed \cite{hansen_fourier_2014} (solid blue). While truncation suppresses internal reflections, it introduces a sharp discontinuity manifested as a rapid amplitude change exceeding \qty{10}{\percent} of the peak amplitude value over the final \qty{1}{\ps} (inset). This violates the smooth-decay criterion for reliable Fourier transformation, distorting both the transmittance and permittivity spectra (Figs.~\ref{fig3}b--d), where the phonon resonance appears artificially narrowed and enhanced in intensity.

In contrast, the Gaussian-windowed waveform (blue curve, Fig.~\ref{fig3}a) eliminates discontinuities through smooth tapering (window standard deviation $\sigma=\num{100}~\textrm{points}\approx\qty{3.33}{\ps}$) while maintaining the main pulse's essential shape and amplitude. This apodization reduces spectral sidelobes while suppressing substrate echoes, yielding permittivity spectra where the resonance near \qty{0.5}{\THz} exhibits physically credible amplitude and linewidth---consistent with PbTe's low-temperature phonon behavior. Although both the truncated and Gaussian-windowed signals yield compatible peak frequencies of \qty{0.50(4)}{\THz} and \qty{0.52(2)}{\THz}, respectively, only the latter provides a realistic amplitude, with the extracted $\epsilon_2$ peak estimated as \num{42(1)e3}, closely matching the values reported in Ref.~\cite{baydin_magnetic_2022}. In contrast, the truncated data yields an overestimated $\epsilon_2$ peak of \num{135(1)e3}, highlighting how inappropriate windowing can distort spectroscopic parameters even when the resonance frequency appears preserved. Critically, these results demonstrate apodization's necessity when signals exhibit pronounced post-peak oscillations, though optimal resonance fidelity requires careful window parameter selection.

\subsection{Optimization of window size for resolution-artifact balance}
\begin{figure}[t]
    \centering
    \includegraphics[width=1\textwidth]{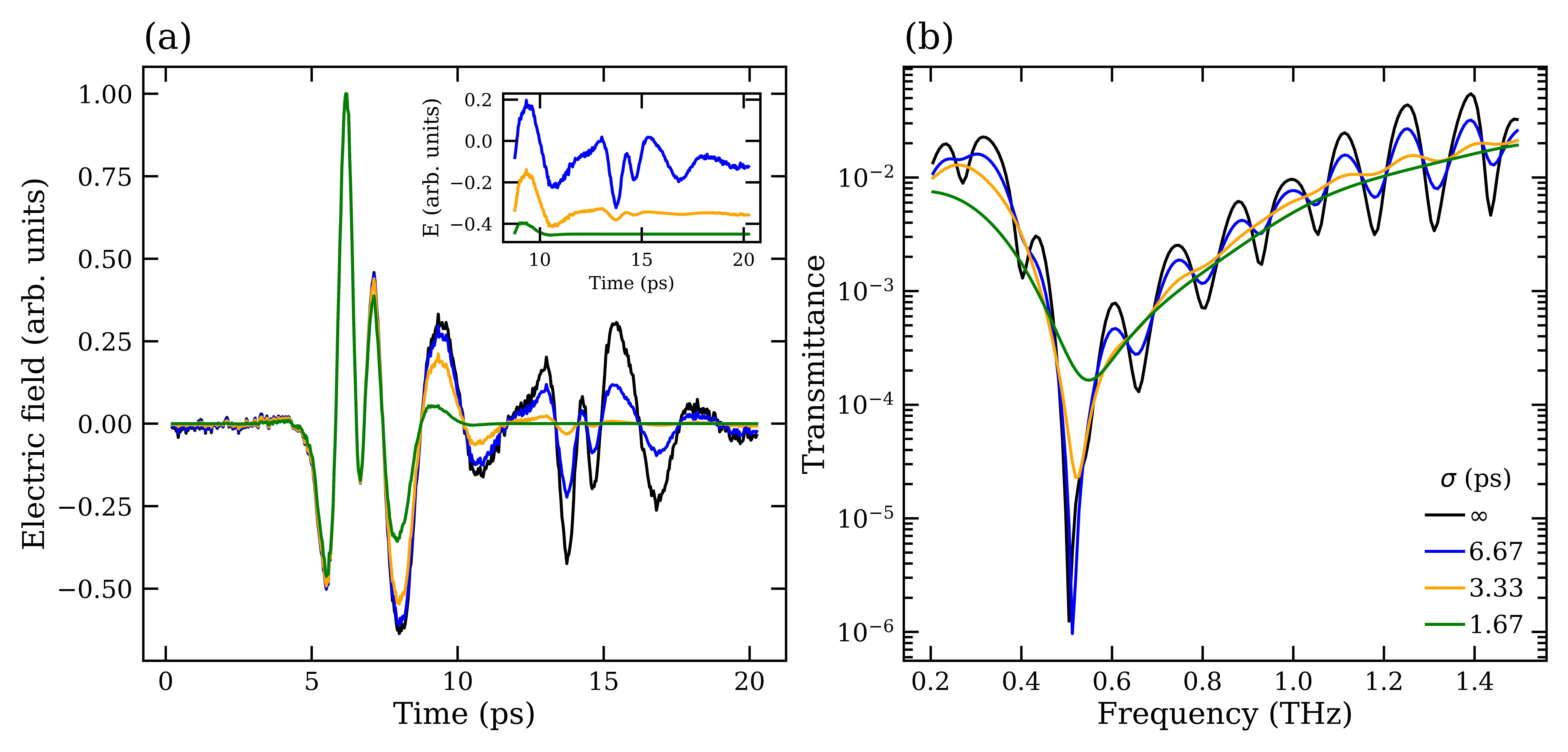} 
    \caption{Effect of Gaussian window length variation on extracted optical properties of a PbTe thin film at \qty{10}{\K}. \textbf{(a)} Time-domain transmitted electric fields after applying Gaussian windows with standard deviations $\sigma=\qty{1.67}{\ps}$ (green), $\sigma=\qty{3.33}{\ps}$ (yellow), and $\sigma=\qty{6.67}{\ps}$ (blue). The black curve shows the original, untruncated signal ($\sigma=\infty$). Inset: vertically-offset zoomed view near Fabry--Pérot echoes highlighting the critical trade-off between echo suppression and signal preservation. \textbf{(b)} Transmittance spectra derived from the windowed signals. The intermediate $\sigma=\qty{3.33}{\ps}$ case (yellow) balances suppression of reflection-induced spectral modulations with preservation of the phonon absorption feature near \qty{0.5}{\THz}, while shorter windows cause artificial broadening/attenuation (green) and longer windows retain spurious oscillations (blue).}
    \label{fig4}
\end{figure}

The window length selection in Step 6 presents a universal trade-off for all apodization functions: shorter windows provide strong suppression of Fabry--Pérot artifacts but risk truncating physically relevant signal dynamics, while longer windows preserve more waveform content at the expense of potentially retaining interference-driven spectral modulations. To address this balance, we systematically vary the window size using the same \qty{10}{\K} PbTe dataset, with Gaussian windowing serving as a representative case in which the standard deviation $\sigma$ directly controls the effective duration.

Fig.~\ref{fig4}a demonstrates that shorter windows ($\sigma=\qty{1.67}{\ps}$, green curve) aggressively attenuate the signal tail, suppressing Fabry--Pérot reflections while unintentionally truncating oscillations encoding the phonon resonance. This compromises the transmittance spectrum (Fig.~\ref{fig4}b) near \qty{0.5}{\THz}, where the absorption feature appears artificially broadened and attenuated in amplitude. Conversely, longer windows ($\sigma=\qty{6.67}{\ps}$, blue curve) preserve nearly all temporal dynamics---closely matching the original signal ($\sigma=\infty$, black curve)---yet retain reflection-induced spectral ripples. The intermediate $\sigma=\qty{3.33}{\ps}$ case (yellow curve) achieves an optimal compromise: sufficient echo suppression while preserving resonance-critical oscillations for accurate lineshape recovery. Although the central phonon frequency remains consistent across window lengths, spectral distortions introduced by longer windows complicate the lineshape and hinder reliable physical modeling, such as Lorentz oscillator fitting.

\subsection{Spectral consequences of window function selection}
Having addressed window length optimization, we now examine the intrinsic influence of window shape on spectral fidelity. Different apodization functions attenuate temporal signals through distinct tapering profiles, yielding characteristic trade-offs between spectral resolution and artifact suppression. These differences are characterized by two fundamental properties: main lobe width (determining effective frequency resolution) and sidelobe suppression (governing attenuation of spurious oscillations from truncation or residual reflections) \cite{harris_use_1978}. In this fourth case example, we systematically evaluate these characteristics' impact on complex permittivity extraction by comparing four established window functions---Gaussian \cite{hansen_fourier_2014}, Barthann \cite{ha_new_1989}, Flattop \cite{dantona_digital_2006}, and Blackman--Harris \cite{harris_use_1978}---applied to the same \qty{10}{\K} PbTe thin-film dataset. Each window length was individually optimized per last section's criteria to balance echo suppression and resonance preservation.

\begin{figure}[t]
    \centering
    \includegraphics[width=1\textwidth]{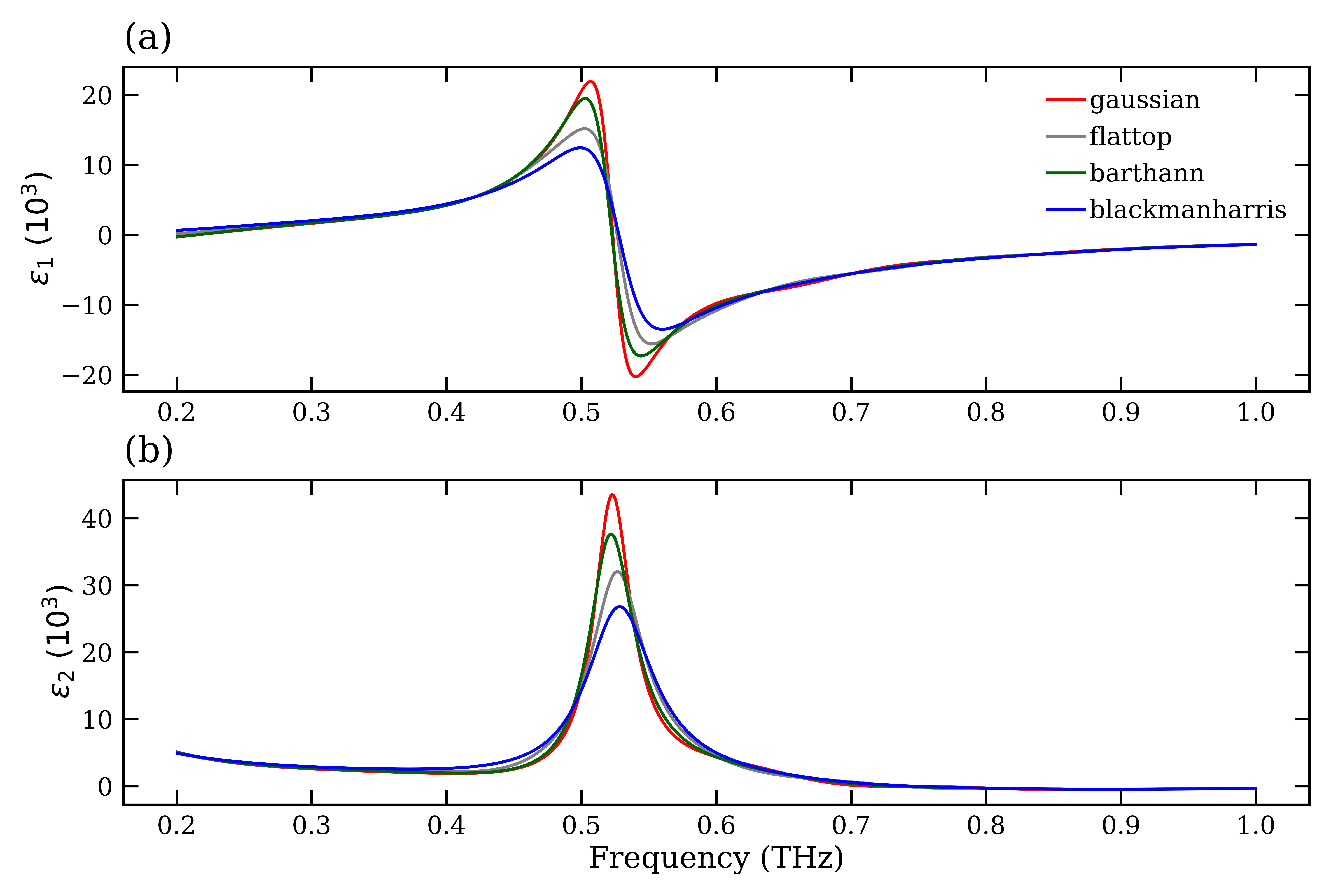}
    \caption{Effect of window function choice on extracted optical properties of a PbTe thin film at \qty{10}{\K}. Each curve corresponds to a different window function applied to identical time-domain data, with window lengths individually optimized for echo suppression and spectral fidelity. \textbf{(a)} Real and \textbf{(b)} imaginary parts of the complex permittivity. Gaussian windowing (red) yields sharper, higher-amplitude features; Flattop (gray) and Blackman--Harris (blue) produce broader lineshapes with attenuated peaks; Barthann (green) exhibits intermediate characteristics.}
    \label{fig5}
\end{figure}

Fig.~\ref{fig5} presents the extracted real and imaginary permittivity components, $\epsilon_1(\nu)$ and $\epsilon_2(\nu)$, obtained from each windowed signal. Substantial differences emerge near the phonon resonance at \qty{0.5}{\THz}, where the Gaussian window (red) yields the sharpest spectral features with maximum amplitudes. This behavior aligns with its narrow main lobe width, which minimizes spectral broadening. Conversely, Flattop (grey) and Blackman--Harris (blue) windows produce significantly smoother spectra with attenuated peak amplitudes and broader lineshapes---consistent with their enhanced sidelobe suppression and wider main lobes. The Barthann (green) window exhibits intermediate characteristics: it preserves greater resonance sharpness than Flattop or Blackman--Harris while showing less pronounced amplitude enhancement than the Gaussian. This comparative analysis demonstrates that no single parameter (such as phonon position) suffices to define optimal performance, rather, window selection should prioritize spectroscopic goals: windows like Gaussian when resolving sharp resonances is critical, versus windows like Flattop or Blackman--Harris when quantifying absolute amplitudes across broad spectral ranges is essential. Intermediate options like Barthann provide balanced alternatives when both priorities must be accommodated.

\section{Discussion}
A central methodological aspect of the routine for extracting complex permittivity from transmission-based THz-TDS of thin films is determining whether simple truncation suffices or if smooth windowing is required. While our approach primarily relies on visual inspection of time-domain signals to identify discontinuities near truncation boundaries---a strategy proven effective across the studied cases---we also incorporated semi-quantitative assessments of slope steepness at truncation points. This motivates developing more systematic strategies based on explicit numerical indicators for future implementations.

To advance this goal, we propose two complementary metrics computed from the truncated electric field waveform $\bar{E}(t)$. The first, $D_1[\bar{E}]$, quantifies the relative amplitude discontinuity at the truncation position $t^\textrm{refl}$:
\begin{equation}
D_1[\bar{E}] = \frac{\left|\bar{E}(t^\textrm{refl})-\bar{E}(t_0)\right|}{\max\left|\bar{E}(t)\right|},
\end{equation}
while the second, $D_2[\bar{E}]$, measures the relative slope change:
\begin{equation}
D_2[\bar{E}] = \frac{\left|\frac{\mathrm{d}\bar{E}}{\mathrm{d}t}(t^\textrm{refl})-\frac{\mathrm{d}\bar{E}}{\mathrm{d}t}(t_0)\right|}{\max\left|\frac{\mathrm{d}\bar{E}}{\mathrm{d}t}(t)\right|}.
\end{equation}
These metrics offer a quantitative foundation for detecting discontinuities that may cause spectral artifacts and could enable algorithmic windowing decisions. However, $D_2[\bar{E}]$ requires careful implementation, given that derivative-based metrics are inherently noise-sensitive and should be coupled with appropriate smoothing or local averaging.

A fundamental consideration in THz-TDS is whether extracted spectral features support reliable physical modeling. For resonant systems, the Lorentz oscillator model \cite{fox_optical_2010}
\begin{equation}
    \epsilon(\nu) = \epsilon_\infty + \frac{(\epsilon_\textrm{s}-\epsilon_\infty)\nu_0^2}{\nu_0^2-\nu^2-i\nu\gamma}
\end{equation}
---where $\epsilon_\infty$ and $\epsilon_\textrm{s}$ are high-frequency and static dielectric constants, $\nu_0$ is the resonance frequency, and $\gamma$ the linewidth---is widely used to characterize dielectric responses. Accurate parameter extraction critically depends on spectral fidelity, which is strongly influenced by time-domain preprocessing choices.

Windowing directly impacts this fidelity: Improper truncation introduces discontinuities causing spectral leakage, while apodization inherently redistributes spectral information---altering resolution and potentially distorting features. As signal processing principles affirm \cite{prabhu_window_2014}, no universally optimal window exists; selection must align with analysis objectives.

Our results demonstrate this principle: different windows produce distinct resolution--amplitude trade-offs that directly affect modeling priorities. Windows like Flattop \cite{dantona_digital_2006} or Blackman--Harris \cite{harris_use_1978}, with enhanced amplitude fidelity, better support determinations of parameters such as $\epsilon_\infty$ and $\epsilon_\textrm{s}$. Conversely, windows like Gaussian \cite{hansen_fourier_2014}, preserving sharp spectral features, enable more precise extraction of $\nu_0$ and $\gamma$.

In summary, while several alternative spectral estimation techniques exist (including parametric models and wavelet-based approaches) Fourier-based analysis remains the most widely adopted method in THz spectroscopy due to its ease of implementation. In this context, strategic time-domain windowing is essential for resolving intrinsic resonances in thin-films, especially when spectroscopic features overlap with substrate-induced artifacts. In this work, we formalize a structured routine that addresses this challenge by first assessing the feasibility of simple truncation---based on the presence of discontinuities near internal reflection boundaries---and then applying appropriate apodization windows when discontinuities threaten spectral fidelity. Through representative case studies, we also explore practical considerations for window design. The window length should scale with the temporal extent of resonance dynamics, while the window shape should be chosen according to the spectroscopic priorities: maximizing frequency resolution or amplitude accuracy. We note that, rather than introducing a new processing paradigm, this work aims to bridge a practical gap in the existing literature by clarifying how and when to apply windowing in common THz-TDS workflows. By doing so, it enables reliable extraction of complex permittivity while preserving the underlying physical features that might otherwise be distorted or obscured.

\backmatter

\bmhead{Author contributions}
N.M.K. and F.G.G.H. conceived the project. E.M. developed and implemented the analysis routine. E.M. and F.L.C. contributed to data handling and spectral analysis. E.M. generated all figures and wrote the initial draft of the manuscript. N.M.K. revised and edited the final version. E.M. and N.M.K. contributed to data interpretation and contextualization. F.G.G.H. supervised the project throughout. All authors discussed the results and commented on the manuscript.

\bmhead{Funding}
This work was supported by the São Paulo Research Foundation (FAPESP), Grants Nos. 2021/12470-8 and 2023/04245-0. E.M. acknowledges financial support from Grant No. 88887.007580/2024-00 of the Coordenação de Aperfeiçoamento de Pessoal de Nível Superior (CAPES). F.L.C. acknowledges support from the Programa Unificado de Bolsas (PUB) of the University of São Paulo. N.M.K. acknowledges support from FAPESP Grant No. 2023/11158-6. F.G.G.H. acknowledges financial support from Grant No. 306550/2023-7 of the National Council for Scientific and Technological Development (CNPq).

\bmhead{Data availability}
All data that support the conclusions of this work are contained within the article. Additional data will be made available upon request.

\section*{Declarations}

\bmhead{Competing Interests}
The authors declare no competing interests.

\bibliography{sn-bibliography}

\end{document}